\documentclass[twocolumn,amsmath,aps,amssymb]{revtex4}
\usepackage{graphicx}
\usepackage{dcolumn}
\usepackage{setspace}
\usepackage{epsfig}
\begin{document}
\title{Two site self consistent method for
       front propagation in reaction-diffusion
       system.}
\author{ Niraj Kumar and Goutam Tripathy }
\address{Institute of Physics, Sachivalaya Marg, Bhubaneswar 751005,
         India}
\begin{abstract}
 We study front propagation in the reaction diffusion 
 process $A\leftrightarrow2A$ on one dimensional lattice 
 with hard core interaction between the particles. We propose
 a two site self consistent method (TSSCM) to make analytic 
 estimates for the front velocity and are in excellent 
 agreement with the simulation results for all parameter
 regimes. We expect that the simplicity of the method will 
 allow one to use this technique for estimating the front 
 velocity in other reaction diffusion processes as well.\\

\end{abstract}
\maketitle{}

Front propagation in the reaction diffusion system is an 
important field of study in nonequilibrium physics. In many
natural phenomena we often encounter propagating fronts separating
different phases\cite{sar}
. Here, in this paper, we study the front dynamics in the 
reaction diffusion system $A\leftrightarrow 2A$ whose mean field
description is given by the following Fisher-Kolmogorove equation
\cite{fisher} for the local density of $A$ particles $\rho(x,t)
$: $\frac{\partial\rho}{\partial t}=D\frac{\partial^2\rho}{\partial
x^2}+k_1\rho-k_2\rho^2$. Here $D$ is the bare diffusion coefficient
of $A$ particle while $k_1$ and $k_2$ are the rates of creation and
annihilation respectively. This equation arises in the macroscopic
description of many processes in natural science and serves as a
generic model to describe front propagation in a system undergoing 
transition from unstable to stable state. The homogeneous steady
states of this equation are $\rho=\frac{k_1}{k_2}$(stable) and 
$\rho=0$(unstable). Hence if we start with an initial condition 
where both the states coexist, the the stable phase invades the 
unstable one with speed $V$ as a travelling wave of the form
: $\rho=\rho(x-Vt)$. Here, such fronts are referred to as $pulled$
in which the leading edge where $\rho<<1$ plays important role 
in describing the front dynamics.  Such leading edge analysis 
gives $V\ge V_0=2\sqrt{k_1D}$. However, for steep enough initial
condition the minimum velocity $V_0$ is selected and it is known
that convergence to $V_0$ takes place with a very slow power 
law \cite{bram1}\cite{ebert}:  $V(t)\sim V_0+\frac{c_1}{t} 
+O(t^\frac{-3}{2})$, where, $c_1<0$.

In this paper, we study the microscopic realization of the 
process $A\leftrightarrow 2A$ on a lattice. The important feature
to be noted here is that the Fisher equation fails to deal with 
internal fluctuations arising due to discrete nature of the 
reacting species, especially when occupancy per site $N$ is small
\cite{derrida}\cite{kessler}. When $N\rightarrow\infty$, the mean
field results are recovered.

Motivated by the velocity selection principle, in which the leading
edge plays an important role, we propose a procedure called
two-site self consistent method(TSSCM). In this method, we describe
the front dynamics by considering the evolution of occupancy at 
only two sites, the first one is the front site while the other 
one is the site just behind it. In other words, in the frame moving 
with the front we study the evolution of occupancy at a site just
behind it. By applying a self consistent approach (explained later) 
in this method we obtain the analytic estimates for the front velocity 
which are in excellent agreement with the simulation results for all
parameter regimes. In 
our simulation we consider a one dimensional lattice composed 
of sites $i=1,2,..L$, $L$ being the size of the lattice. Initially, we
start with left half of the 
lattice filled with $A$ particles while keeping the remaining right
half empty. Each site can either be empty or occupied by maximum 
one particle i.e. hard core exclusion is taken into account. We 
update the system random sequentially where $L$ microscopic moves 
correspond to one Monte Carlo step(MCS). During each update we randomly
select a site and the particle at the site undergoes the following 
microscopic moves: (1) The particle can diffuse to the neighbouring
empty site with rate $D$, (2) The particle can give birth of one 
particle at an empty neighbouring site with rate $\epsilon$
, (3) The particle gets annihilated with rate $W$, if the neighbouring
site is occupied. Due to these microscopic processes the 
rightmost $A$ particle, which is identified as a front, moves and 
we are interested in finding the velocity with which it moves, which is 
given by\cite{pts}:
\begin{eqnarray}{\label{e1}}
 V=\epsilon-\rho_1(W-D)
\end{eqnarray}
Here, $\rho_1$ is the density at the site just behind the front
particle. The above expression for the front velocity can be 
understood by visualising the front particle as a random walker 
moving with rate $\epsilon+D$ in the forward direction and with rate 
$W\rho_1+D(1-\rho_1)$ in the backward direction.

\begin{figure}[h]
\centering
\includegraphics[bb=51 49 235 172,width=9cm]{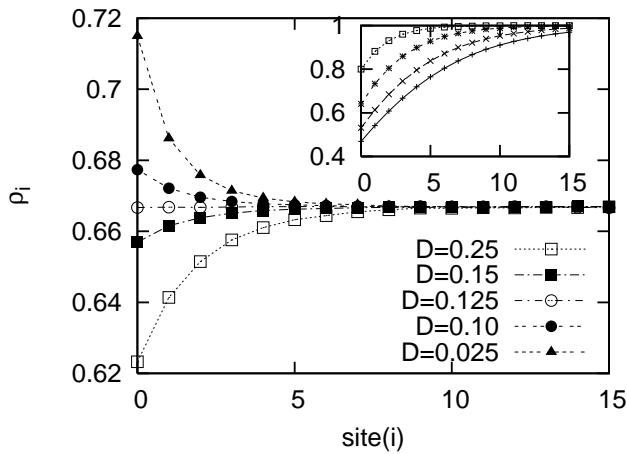}
\caption{Simulation results for the density profile behind the 
front for $\epsilon=0.25, W=0.125$ for different values of $D$
. Inset: density profile for $\epsilon=0.05,W=0$ for $D=0.05,
0.15,0.30,0.45$ from top to bottom. We note that the difference
between $\rho_1$ or $\rho_2$ with bulk density increases with 
increasing magnitude of $|W-D|$ and for $W=D$ all the sites are at
the same bulk density.}
\label{fig:den_profile}
\end{figure}

From Eq.(\ref{e1}), it is obvious that we need to know 
$\rho_1$ in order to predict the front velocity $V$ and there is
no systematic method to find this $\rho_1$ exactly. However, several
approximations have been proposed, for example, in \cite{pts}
, $\rho_1$ was taken as the bulk density $\bar\rho=\frac{\epsilon}
{\epsilon+W}$. This approximation is exact for $W=D$ and shows
reasonable agreement when $|W-D|$ is small but for larger $|W-D|$ we 
observe poor agreement, see Fig.(\ref{fig:den_profile}).
In order to make better estimate for $\rho_1$, especially 
when we are away from the special point $W=D$, in \cite{ng}, we 
proposed a method which systematically improves analytic 
estimate to the desired degree of accuracy. Here we write a 
master equation in the frame moving with the front particle and 
study the evolution of one($l=1$), two($l=2$), three($l=3$) ..etc
sites behind the front. For example, for $l=1$, we study the 
states $\{01,11\}$ and for $l=2$ we have the following set of 
states $\{001,011,101,111\}$, with the rightmost '1' representing
the front. For larger valus of $|W-D|$, we need
to study the states corresponding to larger values of $l$ in 
order to get better analytic estimates but writing the transition
matrix corresponding to the master equation becomes more 
difficult as its size increases as $2^l \times 2^l$. In order to make
analytically tractable approximation for $\rho_1$ and hence the
front velocity we have presented reduced three particle representation.

\begin{figure}[h]
\centering
\includegraphics[bb=51 49 235 172,width=9cm,clip]{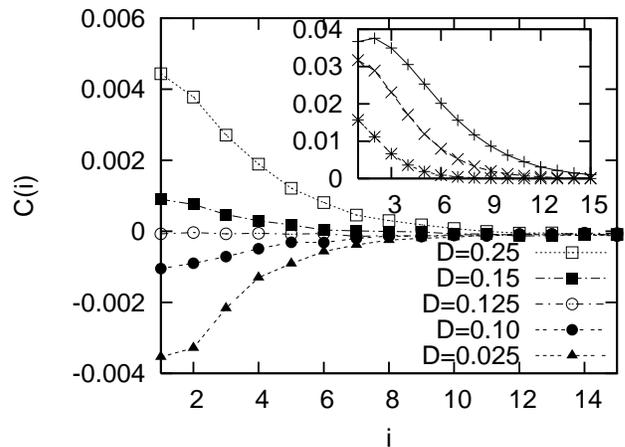}
\caption{Simulation results for the spatial density correlation
between pair of successive sites, $C(i)=<n(i)n(i+1)>-<n(i)><n(i+1)>$, for
$\epsilon=0.25, W=0.125$ for different values of $D$. Inset: 
$C(i)$ versus $i$ for $\epsilon=0.05,W=0$ for $D=0.30, 0.15, 0.05$ from
top to bottom. Here we note that the strength of the correlation 
increases as $|W-D|$ increases.}
\label{fig:den_corr}
\end{figure}

In \cite{ker2}, Kerstein proposed a two particle representation 
for the reaction diffusion process $A\rightarrow2A$, where we study
the evolution of following set of states: $\{11, 101, 1001, 10001,
....\}$, with rightmost '1' representing the front. The 
important point that we note here is that this set of states is 
closed under the transition for $W=0$ while for $W\ne0$ it is not 
closed. This restricts its applicability for nonzero values of 
$W$. Apart from this, it uses the ansatz $p_j=p_0(1-p_0)^j$, where
$p_j$ is the probability that there are $j$ empty sites between the 
first and second particle. This ansatz clearly neglects the spatial 
density correlation which is, in general, not true, see Fig.
(\ref{fig:den_corr}).
In fact this correlation increases with $|W-D|$ and vanishes for the
special point $W=D$. The ansatz also assumes the sites behind the
front to be at the same density $p_0$, which is, of course, not 
true, Fig. (\ref{fig:den_profile}). This leads to a systematic error 
in the analytic estimates which increases gradually with 
increasing value of $|W-D|$. However, one can get better and 
better results by studying the states having larger number of 
particles, for example in \cite{nk1}, we presented mixed  
representation in which we study the following set of states 
$\{011,111, 0101,1101,01001,11001,...\}$ with the rightmost '1' 
representing the front and as expected we get better analytic 
estimates as compared to two particle represenation. But writing
the master equation gets complicated as we increase the number of
particles in the states.

\begin{figure}[h]
\centering
\includegraphics[bb=51 49 235 172,width=9cm,clip]{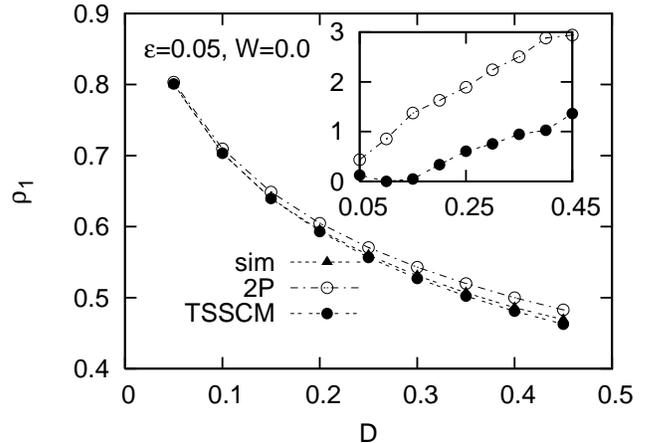}
\caption{Comparision between simulation and analytic results for
$\rho_1$ for $\epsilon=0.05, W=0.0$ and for different values of $D$. The
open circle
corresponds to Kerstein two particle representation
while the closed circle is the result from the present work. We note
that the results of TSSCM are essentially coincident with the 
simulation results. Inset: Percentage relative error in $\rho_1$.}
\label{fig:r1_w0}
\end{figure}

\begin{figure}[h]
\centering
\includegraphics[bb=51 49 235 172,width=9cm,clip]{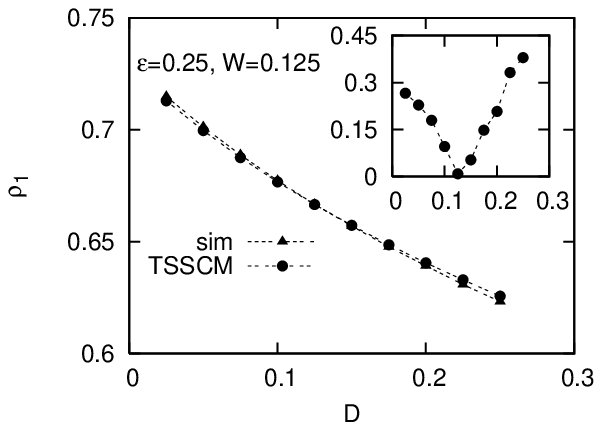}
\caption{Comparision between simulation and analytic results for
$\rho_1$ for $\epsilon=0.25, W=0.125$ and for different values
of $D$. The closed triangle
corresponds to the simulation results while the closed circle is
the results from present work. We note
that the analytic results are essentially coincident with the 
simulation results. Inset: Percentage relative error in $\rho_1$.}
\label{fig:r1_w125}
\end{figure}

\begin{figure}[h]
\centering
\includegraphics[bb=51 49 235 172,width=9cm,clip]{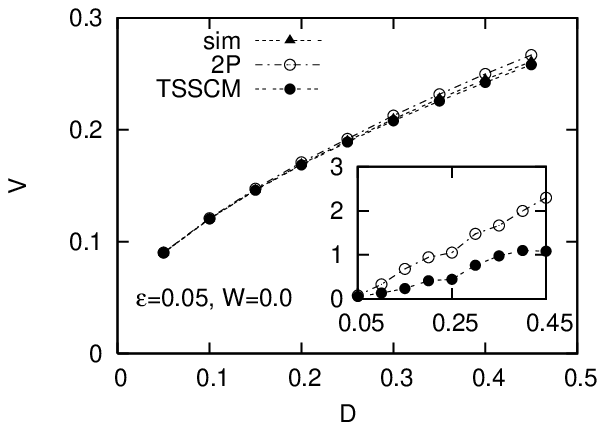}
\caption{Comparision between simulation and analytic results for
front velocity $V$ for $\epsilon=0.05, W=0.0$ and for different 
values of $D$. The open
circle corresponds to Kerstein two particle representation
while the closed circle is the result from present work. We note
that the results of TSSCM are essentially coincident with the 
simulation results. Inset: Percentage relative error in $V$.}
\label{fig:vel_w0}
\end{figure}

\begin{figure}[h]
\centering
\includegraphics[bb=51 49 235 172,width=9cm,clip]{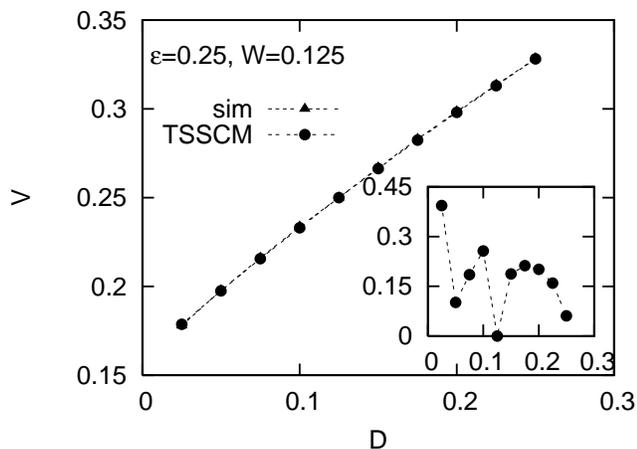}
\caption{Comparision between simulation and analytic results for
 front velocity $V$ for $\epsilon=0.25, W=0.125$ and for different 
 values of $D$ . The 
 closed triangle corresponds to the simulation results while the
 closed circle is the results from the present work. We note
that the analytic results are essentially coincident with the 
simulation results. Inset: Percentage relative error in $V$.}
\label{fig:vel_w125}
\end{figure}

Here, in this paper we present a simple analytic esimate for
$\rho_1$ and hence the front velocity which are in excellent 
agreement with the simulation results. Here we propose a self 
consistent two-site representation, where we write master equation
in the frame moving with the front particle(as in \cite{ng}) for 
the evolution of two states $\{0A, AA\}$. Here the rightmost $A$ 
represents the front particle. In this truncated representation
these two states make transition between each other due to the 
microscopic processes of the model and form a closed set under 
such transitions. For example, $0A\rightarrow AA$ if the leading
particle gives birth to its left empty site with rate 
$\epsilon$. Diffusion of the leading particle to its left changes
the state $0A\rightarrow AA$, provided the second site($F-2$)
behind the leading particle is occupied and remains unchanged if
it is empty. If the probability of occupancy at $F-2$ is denoted as 
$\rho_2$ then the transition $0A\rightarrow AA$ occurs with rate 
$D\rho_2$. Similarly, when the leading particle in the 
realization $AA$ gets annihilated the state of realization changes
to $0A$ if $F-2 $ is empty with rate $W(1-\rho_2)$ while state 
$AA$ remains unchanged if $F-2$ is ocuupied. Considering all such 
transitions one can write the master equation for the state 
probabilities $P_{AA}$ and $P_{AA}$ as,
\begin{eqnarray}{\label{e2}}
\dot P_{0A}&=& (2D-D\rho_2+2W)P_{AA}-(2D\rho_2+2\epsilon+\epsilon
             \rho_2)P_{0A},\nonumber\\
\dot P_{AA}&=& (2D\rho_2+2\epsilon+\epsilon\rho_2)P_{0A}-(2D-D
           \rho_2+2W)P_{AA}.\nonumber\\
\end{eqnarray}
From Eq. (\ref{e2}), it is obvious that we need to know 
$\rho_2$ in order to find the steady state probabilities 
$P_{0A}$ and $P_{AA}$ . In \cite{ng}, $\rho_2$ was approximated as
$\bar\rho$=bulk density =$\frac{\epsilon}{\epsilon+W}$, which works
reasonably well for smaller values for $|W-D|$ but shows poor 
agreement with the larger values of $|W-D|$. We note that 
$\rho_2$, in general, is a function of all the parameters 
($\epsilon$, $D$, $W$) of the model. We express the dependence of 
$\rho_2$ on the parameters as follows:
\begin{eqnarray}{\label{e3}}
\rho_2=aP_{AA}+bP_{AA}^2
\end{eqnarray}
Here, the dependence of $\rho_2$ on the parameters is expressed 
implicitly by $P_{AA}$ whose solution is to be sought and hence 
we call the method self consistent. Now using the fact that when 
$W=D$, $\rho_2=P_{AA}=\bar\rho=\epsilon/(\epsilon+D)$, we rewrite
equation(\ref{e3}) as :
\begin{eqnarray}{\label{e4}}
 \rho_2=(1-\frac{b\epsilon}{\epsilon+D})P_{AA}+bP_{AA}^2
\end{eqnarray}
Using this value of $\rho_2$ in the master equation 
(\ref{e2}) and using the normalisation condition $P_{0A}+P_{AA}
 =1$, we obtain the following equation.
\begin{eqnarray}{\label{e5}}
 \alpha P_{AA}^3+\beta P_{AA}^2+\gamma P_{AA}-\delta=0
\end{eqnarray}
with 
\begin{eqnarray}{\label{e6}}
 \alpha&=&2b\epsilon D+bD^2+b\epsilon^2,\nonumber\\
 \beta&=&D^2+\epsilon^2+2\epsilon D-4b\epsilon D-2bD^2-2b
           \epsilon^2,\nonumber\\
 \gamma&=&2b\epsilon D+\epsilon D+\epsilon^2b+\epsilon^2
            +2\epsilon W+2WD,\nonumber\\
 \delta&=& 2\epsilon^2+2\epsilon D.
\end{eqnarray}
 Now Eq. (\ref{e5}) is in terms of two unkowns $b$ and $P_{AA}$ 
 and hence we must fix $b$ in order to find $P_{AA}$. It is known 
that in the limit $D\rightarrow\infty$, the front moves with the 
Fisher velocity $V_0=2\sqrt{\epsilon D}$. Also from equation 
(\ref{e1}), the front velocity in terms of $\rho_1$ is given as 
$V\sim D\rho_1=DP_{AA}$, when $D$ is very large compared to 
$\epsilon$ and $W$. Equating this velocity with the Fisher 
velocity $V_0$, we get $P_{AA}=2\sqrt{\frac{\epsilon}{D}}$. Now
substituting this value of $P_{AA}$ in Eq. (\ref{e5}), we 
have an equation which is linear in $b$ and which in the limit
$D\rightarrow\infty$ gives $b=\frac{1}{4}$. Substituting this 
value of $b$ in the Eq. (\ref{e5}) we obtain the following
cubic equation in $P_{AA}$.
\begin{eqnarray}{\label{e7}}
 (\epsilon^2&+&D^2+2\epsilon D)P_{AA}^3+(2\epsilon^2+2D^2
  +4\epsilon D)P_{AA}^2
 +(5\epsilon^2\nonumber\\&+&6\epsilon D+8\epsilon W+8WD)P_{AA}
 -8\epsilon^2-8\epsilon D=0
\end{eqnarray}
 One can easily solve the above cubic equation and the density at
 site just behind the front particle $\rho_1=P_{AA}$ can be 
 obtained. The results obtained have been shown in the Figs. 
 (\ref{fig:r1_w0}) and (\ref{fig:r1_w125}) and are in excellent 
 agreement with the simulation results. Here we have also shown
 the percentage relative error in $\rho_1$ i. e., $\frac{|\rho_1^{s}
 -\rho_1^{a}|}{\rho_1^{s}} \times 100$, where $\rho_1^{s}$ and $\rho_1^{a}$
 correspond to simulation and analytic results. Once we know this
 $\rho_1$, the front velocity is obtained from Eq. (\ref{e1})
 and we have shown it in the Fig. (\ref{fig:vel_w0}) and 
 (\ref{fig:vel_w125}). Here, we also observe very good 
 agreement with the simulation results. In Figs. (\ref{fig:r1_w0}) 
 and(\ref{fig:vel_w0}) we have compared the results with that of 
 Kerstein two particle representation. The interesting thing that
 we notice here is that the results obtained from the present 
 work is closer to the simulation results as compared to that
 obtained using two particle representation. 

 To conclude, we have developed a two site self consistent method 
 for the propagating fronts in the reaction diffusion system 
 $A\leftrightarrow2A$ whose results are in excellent agreement 
 with the simulation results for all parameter regimes. We observe that 
 for $W=0$, the results obtained are better than that using Kerstein's
 two-particle representation. We also notice that the present work appears
 to have an advantage over Kerstein's two-particle representation due to two 
 key factors
 : firstly, TSSCM doesnot neglect the spatial density 
 correlation which, indeed, is neglected in two particle representation
 by using the product measure ansatz
 , secondly, TSSCM forms a closed set of states under transitions
 due to the microscopic processes for all parameter regimes while the 
 two-particle
 representation does not provide a closed set for $W\ne0$. The simplicity of 
 our analytic method provides an scope to study the velocity of propagating 
 front in other reaction diffusion processes as well. 
       

\begin{thebibliography}{}
 \bibitem{sar} W. van Saarloos, Phys. Rep. 386, 29 (2003). 
 \bibitem{ker1} A. R. Kerstein J. Stat. Phys. 45, 921 (1986).
 \bibitem{ker2} A. R. Kerstein , J. Stat. Phys. 53,703 (1988).
 \bibitem{pts} D. Panja, G. Tripathy and W. van Saarloos, Phys. 
               Rev. E, 67, 046206 (2003)
 \bibitem{ng} Niraj Kumar and G. Tripathy, cond-mat / 0505599.
 \bibitem{nk1} to be published.
 \bibitem{fisher} R. A. Fisher, Ann. Eugenics , 7, 355 (1937).
 \bibitem{derrida} E. Brunet and B. Derrida, Phys. Rev. E 56, 2597 (1997);
           J. Stat. Phys. 103, 269 (2001).
 \bibitem{kessler} D. A. Kessler, Z. Ner and L. M. Sander, Phys. 
          Rev. E 58,107 (1998).  
 \bibitem{bram1} M. Bramson, Mem. Am. Math. Soc. 44, No. 285 (1983).
 \bibitem{ebert} U. Ebert, W. van. Saarloos, Physica D 146, 1 (2000). 
\end{thebibliography}
\end{document}